%% file: main.tex
\documentclass[12pt, oneside]{article}   
\usepackage[margin = 1in]{geometry} 
\usepackage{graphicx}				
                                
\usepackage[english]{babel}                                
\usepackage{amssymb}
\usepackage{wasysym}
\usepackage{longtable}
\usepackage{textcomp}
\usepackage{topcapt}
\usepackage{booktabs}
\usepackage{dcolumn}
\usepackage{siunitx}
\usepackage{apacite}
\usepackage{tabularx}
\usepackage{url,babel}

\usepackage{setspace}
\usepackage{caption}
\title{How Smart Are `Water Smart Landscapes'?\footnote{This manuscript has been co-authored by UT-Battelle, LLC, under contract DE-AC05-00OR22725 with the US Department of Energy (DOE). The US government retains and the publisher, by accepting the article for publication, acknowledges that the US government retains a nonexclusive, paid-up, irrevocable, worldwide license to publish or reproduce the published form of this manuscript, or allow others to do so, for US government purposes. DOE will provide public access to these results of federally sponsored research in accordance with the DOE Public Access Plan (http://energy.gov/downloads/doe-public-access-plan).}}

\author{Christa Brelsford$^{1}$, Joshua K. Abbott$^{2}$} 
\date{$^1$Geographic Information Science \& Technology Group, Oak Ridge National Laboratory, Oak Ridge TN \\ $^2$School of Sustainability and Center for Environmental Economics and Sustainability Policy, Arizona State University, Tempe AZ}

\begin{document}
\maketitle

\clearpage

\section{ABSTRACT}
Understanding the effectiveness of alternative approaches to water conservation  is crucially important for ensuring the security and reliability of water services for urban residents.  We analyze data from one of the longest-running ``cash for grass'' policies - the Southern Nevada Water Authority's Water Smart Landscapes program, where homeowners are paid to replace grass with xeric landscaping.  We use a twelve year long panel dataset of monthly water consumption records for 300,000 households in Las Vegas, Nevada. Utilizing a panel difference-in-differences approach, we estimate the average water savings per square meter of turf removed.   We find that participation in this program reduced the average treated household's consumption by 18 percent. We find no evidence that water savings degrade as the landscape ages, or that water savings per unit area are influenced by the value of the rebate.  Depending on the assumed time horizon of benefits from turf removal, we find that the WSL program cost the water authority about \$1.62 per thousand gallons of water saved, which compares favorably to alternative means of water conservation or supply augmentation. 

\section{INTRODUCTION} 
Policymakers in many cities and municipal areas are increasingly faced with the harsh reality of water scarcity. 
Drought declarations have become commonplace, with the 2014 Drought State of Emergency in California serving as but one high-profile example. 
This scarcity has been driven by a combination of reduced rainfall and increased demand due to rapid population growth in arid regions such as the U.S. Southwest.
Gaps between water supply and demand were historically addressed by augmenting supply through large scale water infrastructure projects, but now these projects are largely regarded as excessively costly. 
For water utilities, the result has been an increasing focus on fostering water conservation among their customers. 
Economists have frequently advocated for raising water delivery prices to end users since they can allocate the burden of water rationing efficiently across users and encourage customers to target their personal water conservation in its lowest-valued uses first. 
However, raising prices can create undesirable distributional consequences and may be politically unpopular.
As a result, water utilities have tended to favor a range of non-price policies such as watering restrictions, marketing campaigns, and subsidies for modifications to indoor and outdoor water infrastructure \cite{olmstead_comparing_2009}. 

Conservation measures targeting outdoor landscaping have become especially popular, and are often justified on the basis that outdoor water use has constituted 60 to 65\% of residential demand in arid areas over a long time period \cite{mayer_residential_1999,mayer_water_2016}. 
Furthermore, consumers often are often poorly educated about their outdoor water use \cite{attari_perceptions_2014}, suggesting that there may be low hanging fruit for water conservation with even small incentives and changes in customer awareness.
California recently devoted millions of dollars to replace turf with drought friendly landscapes \cite{Goldenstein_guide_2015}.
While the difference in watering requirements of mesic vs. xeric landscaping are well established \cite{mayer_outdoor_2015} and short-run savings have been demonstrated in a few cases \cite{sovocool_-depth_2006,medina_yardx:_2004} a number of questions remain unanswered about turf-removal subsidy programs.
For example, do these programs produce long-term savings, or do they suffer from the offsetting behaviors of the rebound effect exhibited for many energy efficiency interventions \cite{sorrell_empirical_2009,gillingham_energy_2013} and for the installation of low-flow plumbing \cite{campbell_prices_2004} and day-of-week watering restrictions \cite{castledine_free_2014}?\footnote{Aside from offsetting behavior, other causes of falloff in program effectiveness may include leaks from aging drip irrigation and the possibility for increased water demands of vegetation in the face of widespread conversion to xeric landscaping through its effects on outdoor temperatures \cite{klaiber_like_2017, gober_tradeoffs_2012}.} 
Do they conserve water in a cost-effective manner relative to other forms of conservation or supply augmentation? 

To shed some light on these questions, we analyze data from one of the longest-running ``cash for grass'' policies - the Southern Nevada Water Authority's (SNWA) Water Smart Landscapes program (WSL).
This program pays homeowners to replace their lawns with xeric (desert) landscapes.
Utilizing a panel difference-in-differences (DID) approach, we use twelve years of monthly water customer billing data provided by the Las Vegas Valley Water District combined with geocoded spatiotemporal data on program enrollment to estimate the average water savings per square meter of turf removed. To provide a sense of the impact of the program across the year, we estimate water savings separately for four seasons of the year. We also investigate the long-run temporal profile of the program by investigating whether there are discernible differences in water savings across earlier vs. later cohorts of participants and by examining whether water savings attenuate over time. In order to assess the validity of the DID research design we use an event study similar to that employed by \citeA{Davis_cash_2014}. This method allows us to verify that the changes in WSL households' water consumption coincide with their program participation, and to demonstrate that there is no evidence of a contemporaneous consumption change for our control group which may lead to spurious estimates of WSL treatment effects. 
Finally, we estimate annualized water savings per dollar of subsidy spent and compare these costs to estimates of the cost of conserving this water by other means.

\section{BACKGROUND}

\begin{figure}[ht]
   \centering
   \includegraphics[width=4.5in, angle = 90]{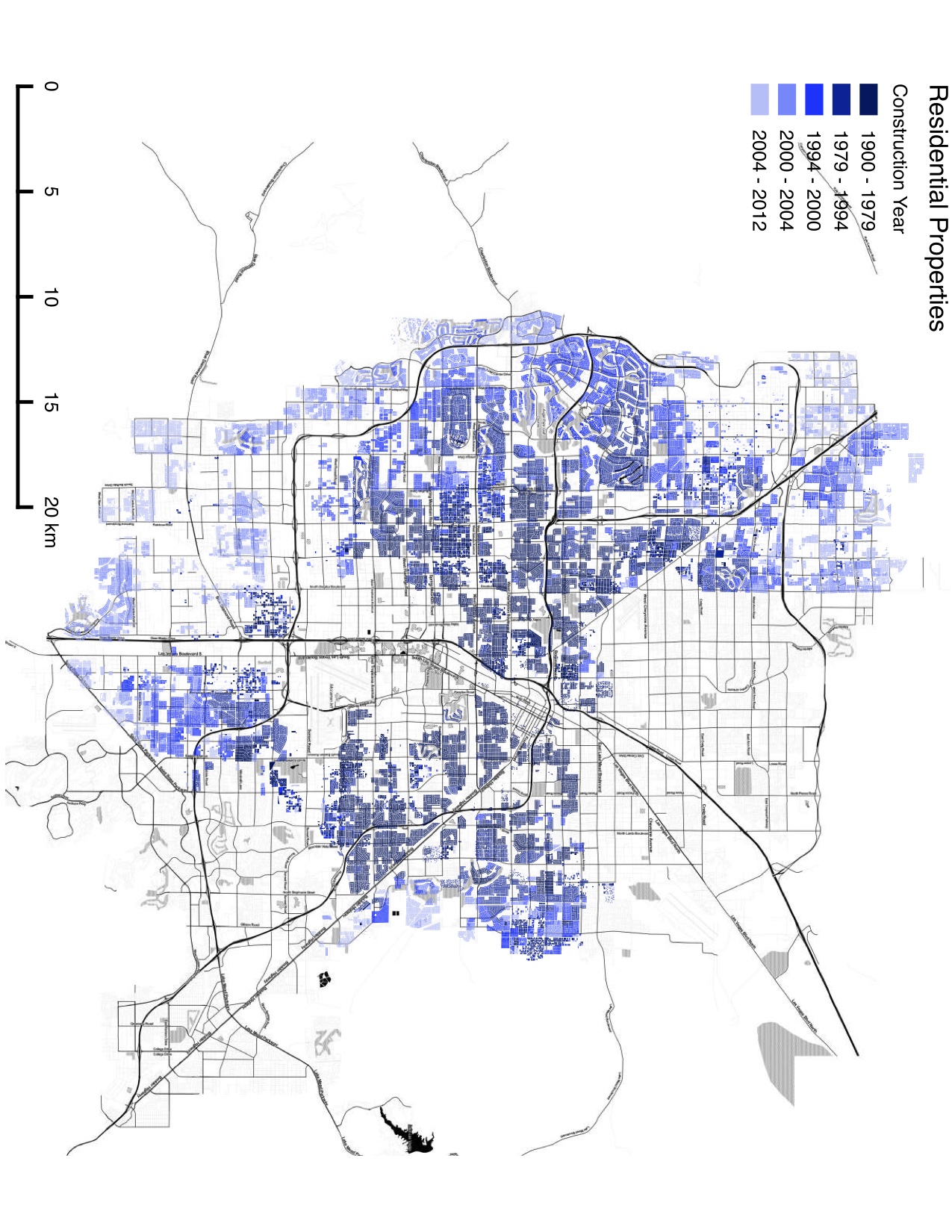} 
   \caption{Residential Parcels inside the study area, which is the urbanized parts of the Las Vegas Valley Water district service area.  Parcels are colored by the in which they were constructed.}
   \label{fig:map}
\end{figure}

\subsection{Water Scarcity in Las Vegas}
Las Vegas, located within Clark County Nevada, has been at the forefront of U.S. ``Sun Belt'' development, with its MSA growing from approximately 850,000 residents in 1990 to nearly 2 million in 2010.
An overwhelming amount of this growth occurred outside of the historical core of the city, with residential land area in the city more than doubling \cite{Brelsford_growing_2017}.
Over 90\% of Clark County's water supply comes from Lake Mead on the Colorado River \cite{snwa_water_2009}. This dependence on a river whose waters are fully allocated and in a multi-decadal drought \cite{castle_groundwater_2014}, combined with Nevada's status as a junior rights-holder under the Colorado River Compact, have heavily shaped the development of Las Vegas' water policy.
The Southern Nevada Water Authority (SNWA) was created in 1991 as a water ``super agency'' comprising five water districts, including the Las Vegas Valley Water District (LVVWD) (which serves approximately three-quarters of Clark County, including all unincorporated areas and the city of Las Vegas) and two sanitation districts.
This body was created in order to cooperatively manage water allocations across its members as well as to coordinate supply augmentation and demand management efforts.
Beginning in the late 1990s, and accelerating with the declaration of a drought alert in 2004, Las Vegas began implementing a range of policies, incentives and building code changes aimed at curbing water use \cite{Brelsford_growing_2017}.
These ranged from rebates for irrigation clocks and pool covers, restrictions on turf in new construction, increases in water prices, and improvements in the ability to enforce and fine residents for conspicuous water waste, such as overspray of sprinklers onto sidewalks. 

Many of the SNWA's conservation efforts focused upon reducing outdoor water use.
This was driven in large part by the fact that Las Vegas receives return flow credits for any water that is withdrawn and subsequently returned to Lake Mead.
Most water used indoors is ultimately treated and returned to Lake Mead such that it is not counted against the SNWA allocation. 
Since a substantial portion of outdoor water use cannot be captured for reuse, reductions in outdoor water use represent a much larger increase in effective supply than an equivalent amount of indoor conservation. 
Perhaps the most well-known of SNWA's efforts at curbing outdoor water use was the Water Smart Landscapes program.  

\subsection{The Water Smart Landscapes Program}

\begin{figure}[htb]
   \centering
   \includegraphics[width=4in]{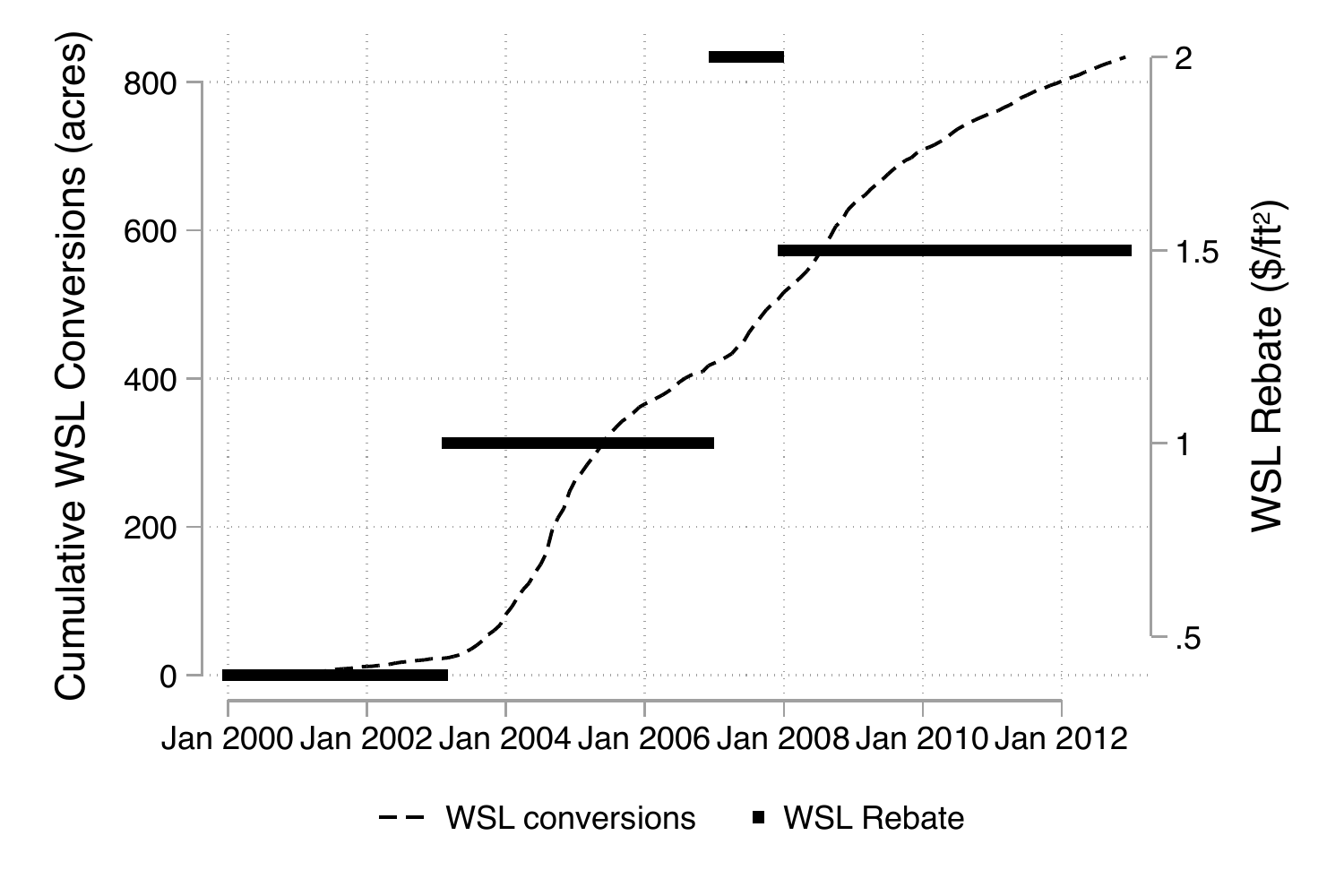}
   \caption{Cumulative WSL conversion area in acres and the nominal WSL rebate at that time. We group WSL participants into four cohort groups based on the nominal rebate price they received. Cohort 1 includes households that participated before February 2003, Cohort 2 includes households that participated between March 2003 and December 2006, Cohort 3 includes January to December 2007 conversions, and Cohort 4 includes households that participated after January 2008. }
   \label{fig:WSL_summary}
\end{figure}

SNWA began the WSL program in 1996 as a small pilot program, and expanded it to all customers in 1998.
Beginning in mid-2000 the program took on its modern form by issuing rebates to customers who converted their lawns to desert landscaping based upon the size of the converted area. 
Fig. \ref{fig:WSL_summary} shows the cumulative area of WSL conversions over the program's history, demonstrating how it grew from a relatively small-scale program to a widespread and important aspect of SNWA's water supply security plan after the 2004 drought declaration.
The program paid residential and commercial landowners between \$0.50 and \$2.00 per square foot of grass removed and replaced with xeric landscaping.  
SNWA notes that typical landscape conversions cost about \$15 per m$^2$ (\$1.40 per square foot in 2000 dollars), although higher end landscapes can cost substantially more \cite{sovocool_-depth_2006,snwa_water_2014}.
This means that for more recent WSL cohorts, the rebate incentive could cover most of the cost of a typical conversion.
There have been limits on the maximum rebate available for residential consumers.  A more detailed history of the WSL rebate structure and limits is outlined by \citeA{Brelsford_Whiskey_2014}.  



The process of WSL conversion consisted of an application followed by a site visit verifying that the property was eligible in terms of meeting minimum conversion requirements and ensuring that the turf was in fact alive and irrigated.  
Upon approval the landowner could replace their lawn with xeric landscaping or artificial turf. 
After a final site visit verifying the size of the conversion and that the post-conversion landscape met the requirement of at least 50\% living ground cover at maturity, the landowner received their payment. 
On average 163 days (median of 142 days) passed between when a homeowner submitted their application and the rebate check was issued, and 4.3\% of conversions took more than a year to complete.\footnote{Email correspondence with SNWA staff members Kent Sovocool and Mitchell Morgan, November 7th, 2016}  

Aside from changes in the subsidy rates over time, the other major change in the program related to restrictions on the length of time owners were required to maintain the conversion. 
Initially there was no such requirement; however, in February 2003 property owners were required to maintain the converted landscape for 5 years. 
In March 2004, this restrictive covenant was extended to last for 10 years, or until the property was sold. Finally, in June 2009, the program was changed again so that the xeric landscape must be maintained in perpetuity, even after the property is sold. 
Despite these requirements, SNWA staff members have no recollection of any systematic efforts to ensure long-term compliance for converted landscapes.\footnote{Conference call with SNWA staff members Kent Sovocool, Morgan Mitchell and Toby Bickmore, April 22nd, 2014} 
Altogether, homeowners in single-family residential properties in the study area had converted about 834 acres of turf by the end of 2012, in comparison to about 35,300 acres of outdoor residential land (48,727 acres total, when including the footprint of homes).  There are also substantial residential areas of the city that were first built with xeric landscaping, in part because of changing preferences, incentives for water smart construction, and building code changes that limited use of turf.  

\begin{table}[htb]
   \centering
   \topcaption{Water consumption and structural characteristics for homes that had a WSL conversion and homes that did not.  For WSL-participating homes, Matched homes, and Random homes, the first row shows pre-treatment average consumption.  For All Non-WSL homes, average consumption is shown across the whole time series. The Random group shows higher average consumption across all seasons than the the entire Non-WSL population does because of Las Vegas' citywide decline in average consumption across the timeseries. }
   \begin{tabular}{l r r r r}
   \toprule
    &   WSL	& Matched  & Random & All Non-WSL \\
    \midrule
    Spring Consumption	&	18.0	&	15.9	&	13.0	&	12.1 \\
Summer Consumption	&	32.4	&	27.0	&	20.7	&	18.9\\
Fall Consumption	&	25.0	&	21.6	&	17.0	&	15.5\\
Winter Consumption	&	12.8	&	12.1	&	10.3	&	9.8\\
\addlinespace[1ex] Indoor Area (m$^2$)	&	196.6	&	195.6	&	185.5	&	185.8\\
Lot Area (m$^2$)	&	830.5	&	813.6	&	634.4	&	638.3\\
Rooms (No.)	&	6.7	&	6.7	&	6.5	&	6.5\\
Pool Ownership	(\%)&	34.1	&	34.3	&	22.5	&	22.4\\
2012 Value (\$)	&	55,238	&	54,899	&	51,080	&	51,233\\
Construction Year (med)	&	1992	&	1992	&	1997	&	1997\\
\addlinespace[0.5ex]  N	&	26,288	&	26,288	&	26,288	&	270,370\\
 	\bottomrule       
 	\end{tabular}
\label{tab:WSL descriptors}
\end{table}

\section{DATA}
The dataset used in this analysis is a twelve year long panel dataset of individual monthly household water consumption records in urban parts of the Las Vegas Valley Water District Service Area.  
Figure \ref{fig:map} shows a map of all households included in the consumption dataset colored by their construction year. 
Out of the 463,658 homes in the Clark County Tax Assessors records and 39,939 households in the WSL conservation program records, 299,872 homes (including 26,376 WSL participants) are in the study area and so have matched records of residential water consumption.  The 13,533 WSL participating households that are not in the study area have similar physical characteristics to the participating homes that are in the study area. 
Each record includes monthly water consumption and the home's structural characteristics as defined by the Clark County Assessors office in 2012. 
Structural characteristics include indoor area, lot size, number of rooms, bathrooms, bedrooms, and plumbing fixtures, as well as the presence or absence of a pool.  
The cleaned consumption dataset excludes 10,196 (3.2\%) homes because they do not have a matching assessors record and thus cannot be geolocated, 30 of which participated in the WSL program.  We also exclude 65 households because the recorded indoor characteristics for the home are physically impossible. Finally, we exclude 1,655 WSL participating households in the study area because they have multiple recorded WSL conversions during the study period; we focus on homes with single conversions for the sake of clean identification.

This leaves a panel dataset with 40,006,271 household/month observations. 
Consumption records are further checked for consistency and validity in three different ways. First, the first month of non-zero water use recorded for each home is excluded as these month's often show unusually high consumption. 
This excludes 299,872 observations.  Second, negative consumption records and the two months prior to a negative record are excluded. 
This excludes 4,347 observations based on negative values alone (some consecutive), and an additional 6,037 based on the two months prior to a negative observation.  
It appears that this negative billing strategy has been used as a billing correction for spuriously high consumption in prior months. 
(The average within-household z score for these pre-negative consumption records is 3.4, substantially higher than the dataset as a whole.)  
Finally, as a guard against extreme outliers, an additional 55,274 observations are excluded because the within-panel z score is greater than five.  

Although there were sometimes caps on the maximum conversion area that could be rebated or the maximum rebate allowed, we always use the actual area of landscape that was converted rather than the landscape area that was eligible for rebate, and the actual rebate received. 
Finally, unless otherwise noted, the nominal dollar values for water bills, water prices, rebate amounts, and any other payments have been deflated to year 2000 dollars.

\subsection{Seasonality}

To provide insight into the temporal footprint of water savings from WSL, we avoid pooling water consumption across distinct seasons of the year into a single regression in favor of estimating distinct regressions for each of four intra-annual seasons, where household water use within each season is averaged across all months within that season. This approach has the advantage of allowing for more flexibility of control than is typically observed in pooled analyses. This approach allows complete independence between estimates of water savings for different seasons while allowing direct comparisons of the different seasons' regressions. 

\begin{figure}[htb]
   \centering
   \includegraphics[width=4in]{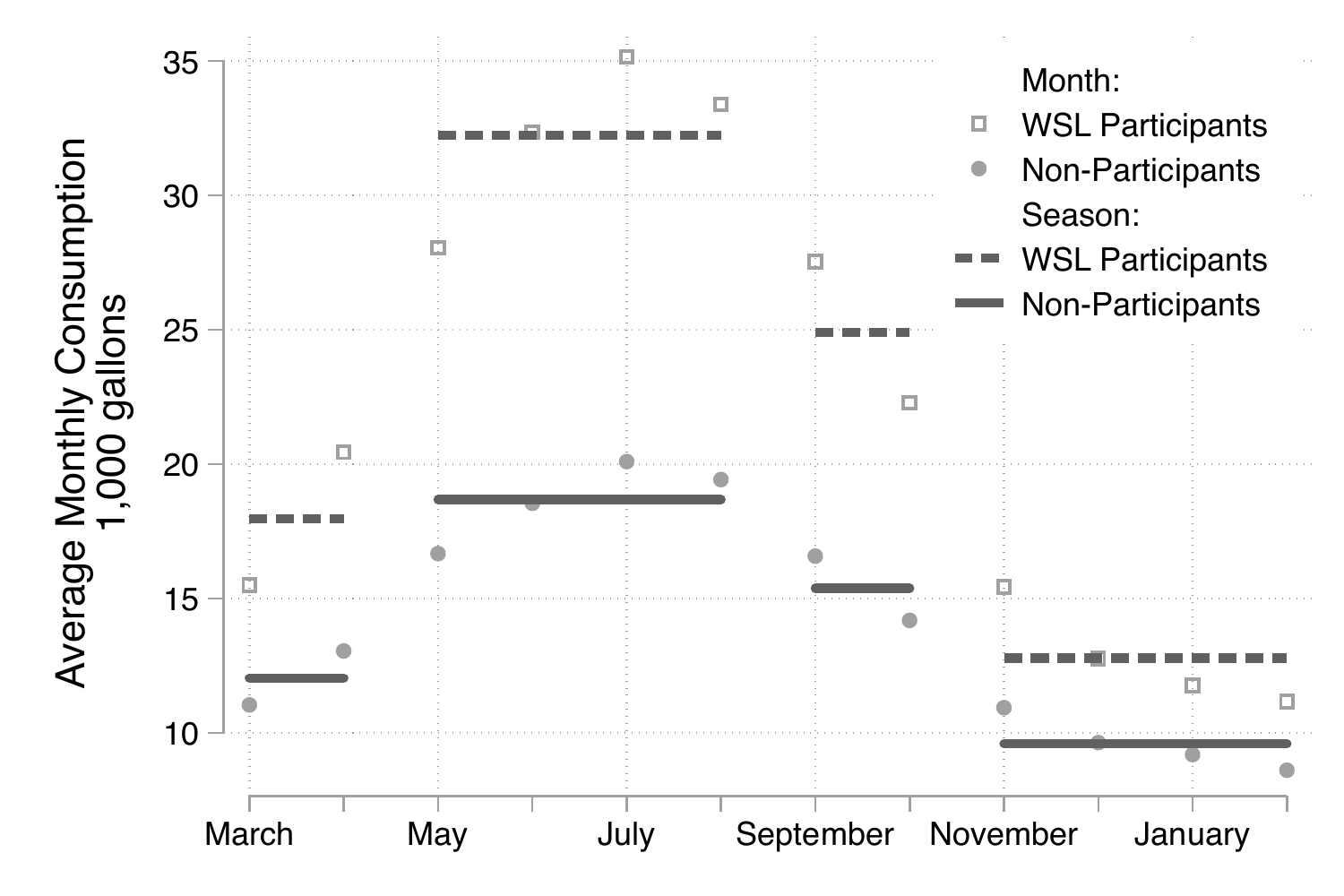}
   \caption{Average Water Consumption by Month and Season, for WSL participating households and non-participating households. Note that WSL participating households consume significantly more water than non-participating households, especially in the summer.  WSL participant data is shown for pre-conversion consumption only.}
   \label{fig:seasonality}
\end{figure}

Our definition of seasons is informed by pooling months with similar water use patterns in Las Vegas' arid desert climate. Spring consumption is composed of the average of March and April consumption.  
Summer is the months of May, June, July and August; and September and October are averaged as fall consumption. 
Finally, November, December, January, and February make up the winter season. 
Since our winter season straddles calendar years, we define the \textit{water year} as running from March to February, where January and February of a given calendar year are included in the previous water year.  
That is, the winter 2004 season's consumption is composed of average consumption from November and December 2004, and January and February 2005.  
Fig. \ref{fig:seasonality} shows average consumption by season and by month for our complete dataset.

Our seasonally-averaged panel dataset consists of about 3.6 million observations across 299,872 households, where 362,849 observations (and 341 households) are excluded based on the criteria described above.  A seasons record is excluded if any one of the monthly records within a season contain flagged data. 
There are 192,654 geolocated households with consumption data in 2000. 
This increases to 297,289 households in 2012 due to Las Vegas' significant population growth and new construction over the intervening years. 

\subsection{Defining Treatment and Control Groups}

For each seasonal regression, the \textit{treatment} group consists of all  households which participated in the WSL program between 2000 and 2012 such that landscape conversion occurred before the end of that season in 2012.  This provides at least one post-conversion datapoint for each household in the treatment group. 

Importantly, households that participated in the WSL program were not generally representative of typical Las Vegas households. 
The most obvious difference, highlighted in Fig. \ref{fig:seasonality}, is that pre-treatment water consumption for WSL households was significantly higher than for the typical household, especially in the summer. 
WSL participating homes also had different structural characteristics (see Tab. \ref{tab:WSL descriptors}): they are older, more valuable, and have larger lot sizes. WSL participating homes also are typically somewhat larger along a variety of indoor dimensions, which is contrary to the overall tendency for older homes to be smaller than newer homes. 

Our panel DID estimation approach fundamentally relies upon the assumption that the counterfactual water use by WSL homes after treatment is adequately imputed from homes that are not yet treated - after controlling for time-invariant differences between these treated and control homes and any shared trends. The substantial differences between WSL and non-WSL homes suggests this ``common trends'' assumption may be questionable. Therefore, we follow \citeA{Ferraro_performance_2014, Ferraro_Panel_2017}
by matching each treatment household with a non-participating household that has a valid consumption record in the year prior to the treatment home's WSL conversion and that has similar location and infrastructure characteristics. 
By selecting a control group that is as similar as possible based on available pre-treatment characteristics that may influence household water consumption, we seek to control for factors that may cause differential trends in residential water consumption between WSL and non-WSL homes. 

The control group consists of an equal number of households which did not participate in the WSL program before 2012. 
The control household selected to match a given treatment household is the consumption record of a home with the most closely matching physical infrastructure characteristics from among all homes in the same block group. 
Control homes are matched on block group (exact), vintage (date) of construction (nearly exact\footnote{Because not all treatment homes have valid consumption data in all seasons of a given year, there are small differences in the number of matching control homes in each season.  
For all but 261 homes there is an exact match for homes of the same vintage in the same block group.  
For the homes for which there is not an exact match, 187 find a match within 1 year, 72 between 2 and 4 years, and twelve homes with differences of vintage beyond 4 years. 
The homes without close matches are all pre-2000 homes, matched with other pre-2000 homes.  
Our results do not change significantly when homes with a poor match on vintage are excluded.}), and the normalized minimum distance for indoor area and lot size.  To find the minimum normalized distance we normalize lot size and indoor area by their respective standard deviations in the WSL population, and calculate the individual z-scores for treatment and candidate control homes.  Then, the distance metric is calculated using the Euclidean norm: $\sqrt{(z^t_{lot}- z^c_{lot})^2 + (z^t_{indoor}- z^c_{indoor})^2}$.  
Some control households are the best available match for multiple treatment households; these households are included multiple times in the control dataset. Table \ref{tab:WSL descriptors} shows that the matched homes are substantially more similar in characteristics that are known to influence water use as well as in pre-treatment consumption levels. 


\section{ESTIMATION APPROACH}

\subsection{Event Study}


To examine the plausibility of the identifying assumptions that underlie our use of the difference-in-differences estimator, we first estimate season-specific event studies of the form shown in Eq. \ref{eq:event_study} using our balanced 
sample of WSL treated homes and matched control homes. 
\begin{equation}
c_{it} =  \alpha_{bWSL(i)} + \sum_{k = -11}^{k = 11} \beta_{k}1[\tau_{it} = k]_{it} +  \sum_{k = -11}^{k = 11} \delta_{k}WSL_{i}\cdot1[\tau_{it} = k]_{it}+ \gamma_t + \epsilon_{it}
\label{eq:event_study}
\end{equation}
where $c_{it}$ is average monthly water consumption for household $i$ in year $t$ over the focal season. For treated homes, \textit{event year} $\tau_{it} = 0$ is defined as the first consumption season in which the effects of a WSL 
conversion are likely to be experienced. Specifically, a treatment household is assigned  $\tau_{it} = 0$ in the first season in which a completed WSL conversion is approved by SNWA, and for each of the three subsequent seasons (i.e. the first year of treatment). 
For example, in a household with a WSL conversion that was finalized in September 2003, $\tau_{it} = 0$ for fall and winter 2003, and spring and summer 2004. In the case of the control properties this time index is set identically to 
the treatment home to which it is matched. Shared absolute time trends across WSL participating homes and their matched controls are captured by water year fixed effects $\gamma_t$.

It is necessary to omit one relative time period as the base category that is absorbed into the model fixed effects. We omit period $\tau_{it}=-1$. The result is that the $\beta_k$ coefficients are interpreted as changes in seasonal water 
consumption relative to the year prior to the landscape change for the matched control group. We also omit $\tau_{it}=-1$ for the WSL-participant homes so that the  $\delta_k$ coefficients reflect deviations in the gap in seasonal water 
consumption between WSL and non-WSL homes relative to the average wedge in the pre-conversion year (captured in the model fixed effects). The model is estimated using fixed effects $\alpha_{bWSL(i)}$ denominated at the Census block-group $b$ and by WSL treatment status (i.e. $WSL_{i}=0,1$) level to control for omitted heterogeneity across space and the treatment and control groups.\footnote{It is not possible to estimate (\ref{eq:event_study}) using parcel fixed 
effects due to the inability to simultaneously identify distinct relative time fixed effects for both treatment and control groups and the absolute time fixed effects using within variation alone \cite {Borusyak_Revisiting_2016}.} Cluster-robust standard errors are used with clusters defined at the block-group level. 

The event study output is useful in several ways. First, it enables us to examine the temporal profile of the $\delta_{k}$ coefficients to test whether the parallel trends assumption between the treatment and control observations is justified on the basis of the pre-treatment data. 
Second, the $\delta_{k}$ reveal the temporal profile of impacts to the treatment group in the time period around WSL conversion - allowing us to assess whether the timing is sensible in light of what is known about the WSL program. 
Finally, by examining the $\beta_{k}$ coefficients for the matched controls and comparing them to the $\delta_{k}$, we can examine whether there are differential trends between the treated and control groups in the pre- and post-treatment periods that are not adequately captured by the water year fixed effects $\gamma_t$. 
The $\beta_{k}$ coefficients also allow us to verify that any treatment effect identified at $
\tau_{it}=0$ is driven by the treatment group itself and not an unspecified shock to the water use of the control homes. 

\subsection{Baseline Models of WSL Effectiveness}\label{sec:methods baseline}

We focus our analysis on the average treatment effect on the treated (ATT) associated with WSL participation in terms of water savings per area of turf removed in gal/m\textsuperscript{2}. 
The focus on a mean \textit{areal} treatment effect, as opposed to a binary indicator of WSL participation, provides a transparent unit of account for WSL effectiveness despite temporal and cross-sectional heterogeneity in the amount of turf removed. 
It also provides a natural metric of comparison since WSL subsidies are denominated by area. 

Prior to presenting our DID specification based upon the available observational data, it is useful to consider the measures of program effectiveness of potential interest to water managers and how these relate to the outcomes of idealized experiments. 
Following \citeA{bennear_municipal_2013} we consider two cases. 
The first, denominated \textit{ATT\textsuperscript{INSTALL}}, is the ATT of a m\textsuperscript{2} of turf removal and landscape replacement 
(i.e. landscape transformation) under WSL among those that participated in the program. This can be envisioned as the outcome of a DID conducted before and after randomized assignment of eventual WSL participants to a treatment 
group that undergoes the landscape transformation and a control group whose landscaping remains unchanged \cite{bennear_municipal_2013}. The treatment in this case is the landscape change itself. 

An alternative measure of effectiveness, denominated \textit{ATT\textsuperscript{SUBSIDY}} is the ATT of having access to WSL subsidy policy (along with the bundled technical advice, certified installers, etc.). This could be estimated using a DID of outcomes from an 
experiment with randomized assignment of households with WSL-eligible yards to a treatment pool with subsidies available and a control pool without these subsidies. An important distinction between these two measures is that the availability of the subsidy need not lead to landscape transformation - both treatment and control households may choose to alter their landscaping as they see fit. 

We are unable to directly estimate either \textit{ATT\textsuperscript{INSTALL}} or \textit{ATT\textsuperscript{SUBSIDY}} using our data. We cannot estimate \textit{ATT\textsuperscript{INSTALL}} because we lack reliable information on the 
landscape changes of non-participants in WSL and therefore cannot guarantee that our control group held their landscaping configuration constant over the study period.\footnote{SNWA does collect limited data on vegetation coverage at the parcel level from remotely sensed imagery.  There are substantial technical challenges to inferring vegetation area estimates in an urban environment from remotely sensed imagery.  These include factors that occlude the image such as clouds or smog, and also trees with large canopies that prevent overhead observation of the ground cover~\cite{Brelsford_using_2014}.  Thus, while we have used vegetation data to check the results, we do not use it as a primary variable.} 
Furthermore, we cannot estimate \textit{ATT\textsuperscript{SUBSIDY}} because the subsidy was made available to \textit{all} eligible households, and we cannot be certain that the adoption rate of water-saving measures (including landscape changes) for any control group we construct will be comparable to that observed absent 
the subsidy. 
Instead, our data allow us to estimate the ATT of WSL participation itself, \textit{ATT\textsuperscript{WSL}}, where both the treatment and control groups experience the same policy environment but the treatment group is 
distinguished by the fact that they enroll in the WSL program. 

\citeA{bennear_municipal_2013} argue in the context of a subsidy program for high-efficiency toilets that $ATT^{SUBSIDY} \leq ATT^{WSL} \leq ATT^{INSTALL}$ (where the measures are framed in terms of absolute values). Parallel
logic applies here. \textit{ATT\textsuperscript{WSL}} is bounded from above by \textit{ATT\textsuperscript{INSTALL}} since the control group for the latter holds landscape constant, whereas it is possible that some individuals in the 
control group for \textit{ATT\textsuperscript{WSL}} adopted water saving landscaping without receiving subsidization. It is also likely that $ATT^{WSL} \geq ATT^{SUBSIDY}$ for the reason that treated individuals in the former case all 
undergo a landscape transformation, while in the latter case those treated with the option of subsidized landscape replacement may choose not to alter their landscape at all. 

We expect that \textit{ATT\textsuperscript{WSL}} is a close approximation of \textit{ATT\textsuperscript{INSTALL}} for the reason that the WSL program was aggressively marketed over much of its history and the 
subsidies under WSL were substantial, covering a substantial portion of the cost of conversion. These factors suggest that the control group for \textit{ATT\textsuperscript{WSL}} should consist primarily of households that chose not to engage in large-scale turf replacement in their yards. 

To develop a baseline estimate of the $ATT$ of WSL participation, \textit{ATT\textsuperscript{WSL}}, we estimate the following regression separately for each of the four previously defined seasons: 

\begin{equation}
c_{it} =  \alpha_i + \gamma_t + \beta a_{it} + \epsilon_{it}
\label{eq:baseline}
\end{equation}
where $c_{it}$ is average monthly water consumption (in gallons) over the focal season in year $t$, and $a_{it}$ is the WSL conversion area (in m$^2$) for each home/year combination. $\alpha_i$ is a parcel-level fixed effect 
reflecting time-invariant unobserved heterogeneity in water use across households which may be correlated with an individual's decision to enroll in WSL. 
The WSL area, $a_{it}$, estimate is proportionally adjusted in any season in which a WSL conversion occurs mid-season.  For example, if a WSL conversion was in place for only two of the four months in a given season, the WSL 
area in that season is adjusted to half of it's true value.\footnote{Alternative approaches, including dropping observations where mid-season conversions occurred yielded very similar results.}  

We estimate Eq. \ref{eq:baseline} using the fixed effects (within) estimator. In order to address problems of serial autocorrelation in individual water consumption \cite{Bertrand_how_2004} as well as spatial correlation in water consumption among neighbors and heteroskedasticity, we report cluster-robust standard errors~\cite{cameron_practitioners_2015}, with clusters defined at the US Census block group level.

\subsection{Durability and Cohort Effects} \label{sec:cohort}

To go beyond the average effects estimated in Eq. \ref{eq:baseline} we test the hypothesis that water savings associated with WSL programs decline as the landscape ages.  
This could be driven by substitution toward other water-intensive uses (e.g., greater indoor water usage) in response to reduced water bills from outdoor watering, increased water needs of maturing vegetation, or gradual degradation of irrigation infrastructure. A simple test of these possibilities can be constructed using the following regression:

\begin{equation}
c_{it} =  \alpha_{i} + \gamma_t  + \beta_0 a_{it} +\beta_{1}  a_{it}  y_{it} +  \epsilon_{it}
\label{eq:rebound}
\end{equation}
where $a_{it}$ is the landscaping area for household $i$ in time $t$ as in Eq. \ref{eq:baseline} that has been converted under the WSL program, and $y_{it}$ is the age, in years, of the WSL conversion. 
Results where landscape-age dummy variables instead of a single slope are used to estimate the effect of the age of the WSL landscape on consumption support the linear approximation used here.\footnote{The WSL conversion age is defined as zero for all household/year combinations before the WSL conversion occurs (rather than as a negative number) and for households which do not ever participate in the WSL program.  
The definition has no practical importance because in that case, $a_{it}$ is also equal to zero, and so the $\beta_0$ terms drop out.}

One factor that could bias our estimates of the durability of WSL water savings is the potential for heterogeneous ATT estimates across different phases of WSL (see Fig. \ref{fig:WSL_summary}). This potential bias occurs because our measurements of the savings of WSL-converted landscapes at different ages ultimately reflect the proportions of different WSL cohorts in each age. Because of the unbalanced weightings of the cohorts across different age groups, if there are heterogeneous treatment effects across cohorts, their effects may be absorbed in the estimates of the landscape age effect if cohort-specific effects are not controlled for. For example, $\beta_1$ in Eq. \ref{eq:rebound} may suggest a significant attrition of water savings over time if, for example, the late adopters under WSL, which would be disproportionately young landscapes in our dataset, had higher water savings - even if water savings for individual cohorts are permanent. 

In order to simultaneously test for the existence of both duarability and cohort effects, we augment Eq. \ref{eq:rebound} with cohort-specific estimates of WSL area at age zero, while still maintaining the interaction of age and WSL conversion area: 

\begin{equation}
c_{it} =  \alpha_{i} + \gamma_t  + \beta_0 a_{it}  +\sum_{j = 1}^{3}  \beta_{j} d_{ji} a_{it} +  \beta_{4}  a_{it}  y_{it} + \epsilon_{it}
\label{eq:rebound cohort}
\end{equation}
where $d_{ji}$ is is a dummy variable that is equal to 1 when household $i$ is in cohort $j$ and 0 otherwise. The cohorts are defined in Fig. \ref{fig:WSL_summary}, and coincide with changes in the marginal rebate value. 
We use the most recent cohort group (group 4, conversions occurring after 2008) as the base level, represented in $\beta_0$.
Then, $\beta_{j}$ for $j \in [1,3]$  can be interpreted as the differential effectiveness for cohort j in comparison to the base group.
Finally, $\beta_4$ is the estimate of any effects of age on the permanence of WSL water savings, after controlling for heterogeneity across participation cohorts.


\subsection{Economic Analysis}

We explore the economic case for the WSL program from both public and private perspectives. From the public perspective, we consider the cost-effectiveness of WSL in terms of the water savings induced per dollar spent. We focus on cost effectiveness rather than employing a full-fledged benefit cost analysis due to the difficulties of estimating the social cost of water for Las Vegas. Furthermore, for much of the period of our analysis Las Vegas has been compelled by drought-induced scarcity to find immediate means to enhance return flows to Lake Mead. Therefore cost-effectiveness seems appropriate for the decision context.

From a private perspective, we estimate the annualized benefits to residents from WSL in terms of lower water bills and compare the stream of these benefits to the costs associated with the landscape conversion. We use this comparison to examine the strength of private incentives for turf removal in the absence of subsidization - providing evidence for whether WSL primarily rewarded landscape conversions that were likely to have occurred even in the absence of the incentive vs. inducing conversions that would not otherwise have occurred (i.e., additionality). 

\subsubsection{Cost-effectiveness}

In order to provide a consistent and interpretable estimate of the water savings that the WSL rebate payments generated, we need to define a projected lifespan for the associated water savings and also develop a temporally consistent method of comparing the water savings to the rebate payments. WSL rebates are given as an upfront payment for water savings that accrue over a long period of time.  In order to resolve these temporal scales, we calculate the annuitized cost of providing the subsidy - effectively the ongoing monthly cost of the debt associated with raising this one-time rebate payment, hereafter referred to as the annuitized subsidy payment, $P_{it}$.\footnote{While SNWA paid WSL rebates out of its regular operating budget, they did issue bonds over our study period and therefore we consider the opportunity cost of budgetary resources to be defined by the cost of capital.} 
 
We must also consider that the water savings from WSL should not be attributed to a parcel indefinitely; eventually many homeowners may have converted to water-saving landscaping without subsidization. Furthermore, in the absence of incentive-based programs like WSL, more draconian emergency policy measures may have been necessary to achieve water conservation goals, inducing otherwise hesitant homeowners to install a xeric landscape. Therefore, the water savings of WSL (and hence the annuitized costs of securing them) should be calculated over the expected term until the landscape would have transitioned to xeric cover in the absence of the subsidy. There is no defensible single estimate of this term, and so we consider durations of 5, 10, 20, and 40 years. To calculate the annuitized cost we utilize the real cost of capital for the SNWA as reflected in the coupon rates of municipal bonds issued by SNWA and Las Vegas in the mid-2000s.\footnote{Nominal rates on municipal bonds issues by SNWA and Las Vegas averaged approximately 5\% in the mid-2000s. The annual real cost of capital is 2.36\% after adjusting for a mean inflation rate of 2.58\%. The equation used to calculate the annuitized subsidy payment is $P_{it}=\frac{r\cdot L_i}{1-(1+r)^{-12n}}$, where $r$ is the monthly real cost of capital, $n$ is the term length (in years), and $L_i$ is the lump sum subsidy payment.} 


Using the annuitized subsidy payment, we estimate panel DID models analogous to Eq. \ref{eq:baseline}:

\begin{equation}
c_{it} =  \alpha_i + \gamma_t + \beta P_{it} + \epsilon_{it}
\label{eq:cost effect baseline}
\end{equation}
Thus, $\beta$ can be interpreted as the marginal monthly water savings associated with an additional monthly dollar spent on WSL rebates. 
The longevity of the WSL program as well as heterogeneity in subsidy payments and water savings across cohorts suggest that the cost effectiveness of WSL has likely varied over time. Thus, we also estimate a model analogous to Eq. \ref{eq:rebound cohort}:

\begin{equation}
c_{it} =  \alpha_{i} + \gamma_t  + \beta_{0} P_{it}  +\sum_{j = 1}^{3}  \beta_{j} d_{ji} P_{it} +  \beta_{4}  P_{it}  y_{it} + \epsilon_{it}.
\label{eq:cost effect cohort}
\end{equation}

\subsubsection{Private Benefits}

In order to estimate the private benefits households receive from participating in the WSL program in the form of reduced water bills, we estimate seasonal regressions of the form shown in Eq. \ref{eq:baseline}, where the dependent variable, mean seasonal consumption, is replaced with the mean seasonal water bill.

\begin{equation}
B_{it} =  \alpha_i + \gamma_t + \beta a_{it} + \epsilon_{it}
\label{eq:private baseline}
\end{equation}
where $B_{it}$ is the monthly water bill for household $it$ in deflated (2000) dollars.  
The $\beta$ coefficient from this regression provides estimates of the average monthly reduction in the water bill in each season per m$^2$ of turf removed. By comparing these estimated water savings to the typical average cost per m$^2$ of removing turf and re-landscaping, we are able to assess whether investing in WSL-style landscapes is economically sensible from a private perspective in the absence of subsidies and for reasonable discount rates.\footnote{We do not consider whether there is any differential positive or negative amenity value to homeowners from the landscape itself. This could be assessed using hedonic price models; however, any amenity value must be considered apart from any capitalized water savings (or potential increases in energy bills) from the xeric landscaping. \citeA{klaiber_like_2017} find evidence in Phoenix, AZ that mesic landscapes have a higher implicit value to homeowners than xeric landscapes, even after controlling for neighborhood microclimate.} 


\section{RESULTS}

As a whole, we find clear evidence that the savings generated by the WSL program are consistent with (though somewhat smaller) than previous engineering estimates by \citeA{sovocool_-depth_2006} and previously published estimates relying on neighborhood average consumption rather than household consumption \cite{Brelsford_growing_2017}.

\subsection{Event Study}

\begin{figure}[ht]
   \centering
   \includegraphics[height = 3.6in]{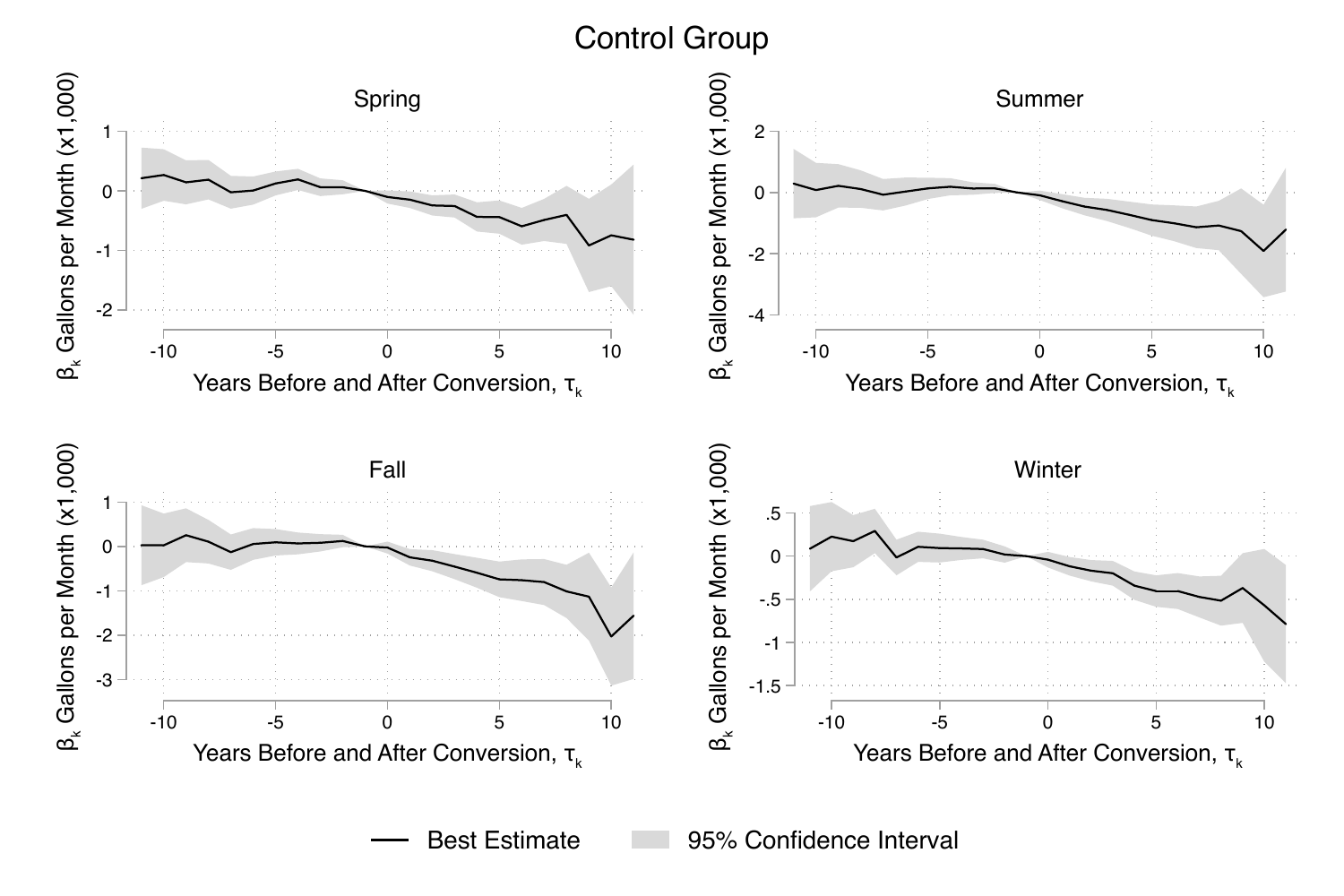} 
   \caption{Event Study results by season for the control group of households that did not participate in the WSL program and are matched on block group, home vintage, lot size and indoor area.}
   \label{fig:es_control}
\end{figure}

Fig. \ref{fig:es_control} shows the $\beta_k$ coefficients for $k \in[-11,11]$ for the matched control group in each season. The lack of significant pre-treatment trends for the control group suggests that baseline trends are well accounted for through the use of the $\gamma_t$ fixed effects. Furthermore, there is no evidence of any shock to the matched control group coinciding with the timing of WSL, with $\beta_0\approx0$ in all seasons. This suggests that any effect recovered by the DID estimator based upon water use patterns in the period immediately after WSL adoption will be linked to changes in the water use of WSL homes. However there is a relatively mild downward trend in water use among the controls in the post-treatment period, even after controlling for shared absolute time trends between WSL and non-WSL homes. This suggests a potential need to control for differential trends in the DID model \cite{Davis_cash_2014}.   

\begin{figure}[ht]
   \centering
   \includegraphics[height = 3.6in]{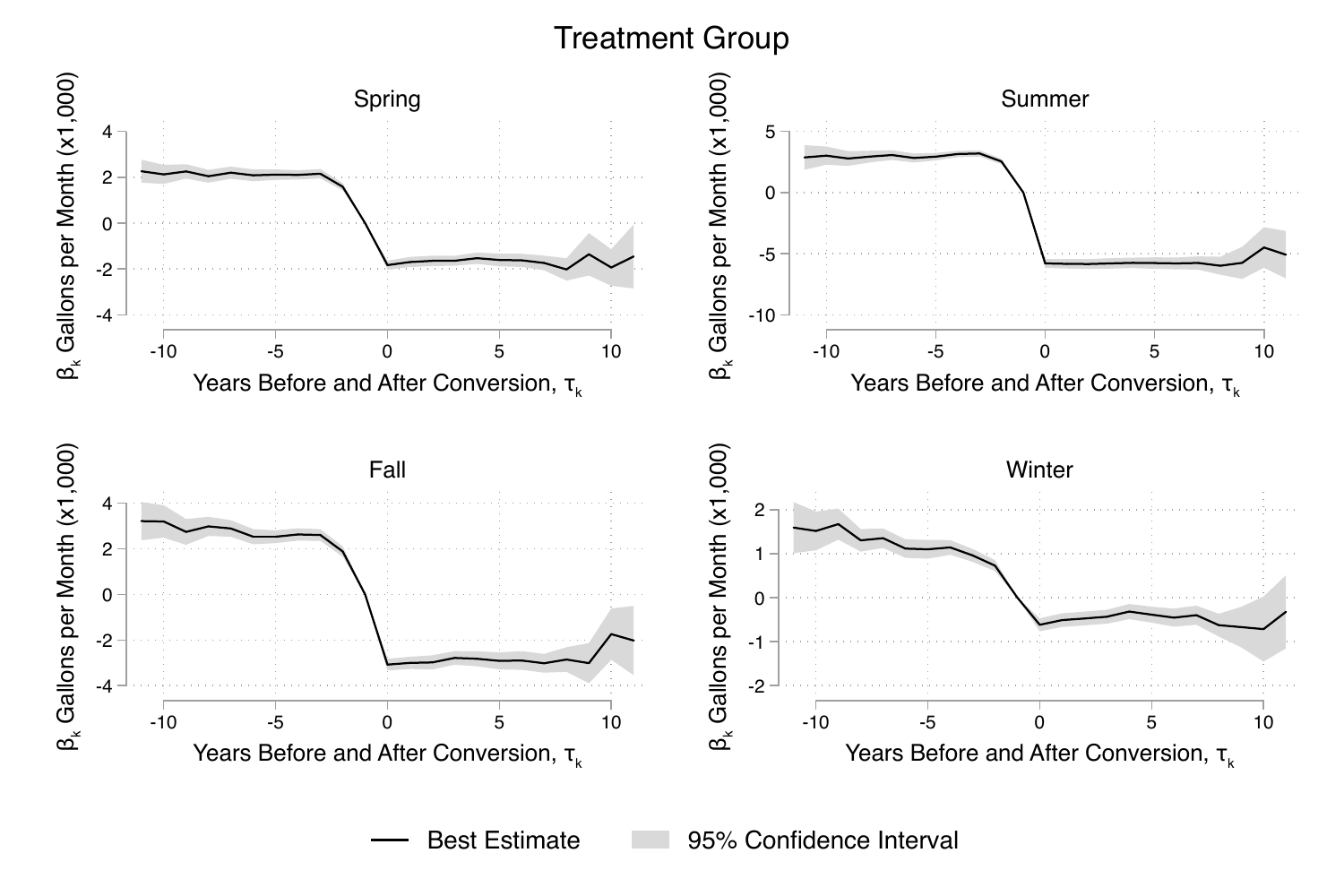}
   \caption{Event Study Results by season for WSL Participants.}
   \label{fig:es_treatment}
\end{figure}

Fig. \ref{fig:es_treatment} shows the $\delta_k$ coefficients for the treatment group in each season - the wedge between the mean water conservation of treatment and control groups relative to its value in the year before WSL conversion ($\tau=-1$), where this differential is normalized to zero. 
The fairly flat pattern of the estimates up to two years prior to WSL installation shows that the difference in water useage between treated and non-treated homes did not change with time prior to the WSL conversion - supporting the common trends assumption underlying the DID estimate.  
All of the reductions in water use for treated households occurred between $\tau = -3$ and $\tau = 0$.  
These differences stop accumulating once the landscape has been fully installed, but begin to appear a couple of years prior to the final WSL inspection. 
In spring, summer and fall, we see that 5 to 15 percent of the decline in water use relative to the control group occurs between two and three years prior to conversion approval; 30 to 40 percent occurs between one and two years prior to final approval; and 45 to 65 percent of the total declines occur in the year prior to conversion. 
After the sharp drop in water use WSL homes show a remarkably flat and persistent effect, with no notable post-treatment trend relative to the $\beta_k$s shared with the control group. 

The reasons for the anticipatory declines in water use prior to final WSL conversion are not observable from the data. 
However, a significant portion of the gap may be driven by the interval between the timing of physical landscape replacement and measured final approval.
Our dataset only includes the date on which a household's WSL conversion was formally approved by SNWA. We do not have any household level data on when the actual landscape change was implemented, even though we were able to obtain some aggregate data on the time elapsed between when a potential participant first contacted SNWA about participating, and the final approval date.
Based on the terms of the WSL program, the actual landscape change must have occurred between the time the participant first contacted SNWA and the approval date that we record in our dataset.

Recall that we define $\tau = 0$ to be the first year in which the landscape has been approved by SNWA, and thus we can be certain that the WSL landscape had been fully installed.  
Because of the annual structure of our dataset, a converted landscape may have been approved up to twelve months prior to the end of the season in which $\tau = 0$.  
Summary data provided by SNWA show a median gap of 4.75 months between the initiation of the conversion process and final approval, with four percent of households taking longer than a year.  
To give a rough sense of how these data structures combine with the gap between landscape change and approval (and consequently the lag between landscape change and when we define $\tau = 0$), we consider the share of households that could have been living with a converted landscape when we define $\tau = -1$. 
If all conversions had a 4.75 month gap between landscape change and SNWA approval and conversions were evenly distributed throughout the calendar year, then nearly 40\% of treated households may have been living with a recently installed xeric landscape when $\tau=-1$ and so their water consumption in this time also reflects the changed landscape.
This suggests that a substantial portion of the pre-treatment declines that we observe are attributable to this lag between landscape change and approval.  
For this reason, we exclude $\tau = -1$ from the main dataset used for our DID models except where otherwise noted. 

An alternative reason for the anticipatory declines in water use could be reductions in outdoor water use in anticipation of turf removal.  However, SNWA requires that a lawn be alive at the time of it's removal for the conversion to eligible for a rebate, so withholding water from a landscape entirely is not likely. Nevertheless, homeowners could partially reduce water consumption in anticipation of the landscape change.
Alternatively, some homeowners may have made other investments in water efficient appliances or fixtures or adopted water-conserving behaviors in advance of their WSL installation. To the extent that WSL-adopters were more likely to make these investments relative to control households, they may be a source of upward bias in estimated water savings from WSL. However, if most of the early reductions in water use represent anticipatory reductions in water use ultimately tied to WSL, then the estimation strategy presented in Eq. \ref{eq:baseline} may understate the effects of WSL due to some reductions in water use occurring before the measured treatment date.  We are unable to specifically differentiate these alternative hypotheses from our data, but examine the sensitivity of our estimates to alternative assumptions below.

Broadly speaking, the magnitude of the estimated WSL-induced  reductions in water use across seasons is consistent with expectations from seasonal differences in water consumption and vegetative water needs.
In the spring, we observe a drop in consumption of about 3,980 gallons/month, summer 8,980, fall 5,680, and in the winter, a decline of 1,570 gallons/month.  
Given that the average size of a WSL conversion is 111 m$^2$, this suggests a reduction in consumption of 36, 81, 51, and 14 gallons/month per converted m$^2$ in the four seasons.

\subsection{Baseline Estimation}

Table \ref{tab:baseline} displays the $\beta$ coefficients and summary statistics from estimating Eq. \ref{eq:baseline}, where each column consolidates the results from four seasonal regressions. To assess the robustness of our 
results, we estimate Eq. \ref{eq:baseline} using a variety of alternative conceptions of the appropriate control group and model specification. Models 1-3 all estimate Eq. \ref{eq:baseline} exactly as specified but vary the definition of the 
control group used to estimate the underlying counterfactual of water use in the absence of WSL adoption. Model 1 randomly matches the water use history of each treated parcel with that of a parcel that never participated in the WSL 
program. Model 2 forgoes the use of a separate WSL-untreated control group altogether, instead utilizing the within-panel variation in the timing of WSL adoption among eventual adopters to identify $\beta$. Model 3 utilizes the matched control group based on 
block group and pre-treatment infrastructure characteristics. 

\begin{table}[ht] 
\centering
\def\sym#1{\ifmmode^{#1}\else\(^{#1}\)\fi}
\topcaption{Regression results for Models 1-5.}
\begin{tabular}{lr rrr rr}
\toprule
	&		&	1	&	2	&	3	&	4	&	5		\\
	\midrule
\input{baseline_1.tex}
\midrule
\multicolumn{2}{l}{Random Control Group}	&	Yes	&	No	&	No	&	No	&	No		\\
\multicolumn{2}{l}{No Control Group}		&	No	&	Yes	&	No	&	No	&	No	\\
\multicolumn{2}{l}{Matched Control} 		&	No	&	No	&	Yes	&	Yes	&	Yes	 \\
\multicolumn{2}{l}{Quadratic Trend}		&	No	&	No	&	No	&	Yes	&	No	 \\
\multicolumn{2}{l}{Include $\tau = -1$ }	&	No	&	No	&	No	&	No 	&	Yes	 \\
\midrule
\input{baseline_2.tex}
\rule{0pt}{3ex}
\input{baseline_3.tex}
\rule{0pt}{3ex}
\input{baseline_4.tex}
\bottomrule
\multicolumn{5}{l}{\footnotesize Standard errors in parentheses}\\
\multicolumn{5}{l}{\footnotesize \sym{*} \(p<0.05\), \sym{**} \(p<0.01\), \sym{***} \(p<0.001\)}\\
\label{tab:baseline}
\end{tabular}%
\end{table}

Models 1-3 attenuate slightly in a consistent manner for all seasons as the control group changes. If our preferred specification, Model 3, is taken as the most reliable estimator, Model 1 overestimates the water savings of WSL by between three and five percent.
Model 2 overstates water savings by an even smaller margin of one to three percent, suggesting that eventual WSL homes provide an effective control group for those that have already selected into the program.   

Model 4 additionally includes a linear and quadratic time trend for the treated group only, based on the same matched sample used in Model 3. This model investigates the possibility, noted within the event study, that differential post-treatment trends between the control and treatment groups may influence our estimates.\footnote{Models utilizing linear trends and higher-degree polynomials generated very similar results.} These results are effectively unchanged from Model 3, showing that despite the small post-treatment decline in our control group shown in Fig. \ref{fig:es_control} our estimates are not driven by a significant differential time trend.  


Model 5 addresses the finding from the event study of declines in water conservation among WSL adopters up to three years prior to confirmation of WSL installation, and our choice to exclude $\tau = -1$ from the preferred specification. 
To assess the ramifications of this phenomenon, we estimate Model 3 but include observations from the year before WSL adoption ($\tau=-1$) for both treatment and control groups. 
The result is an eleven percent decrease in the estimated annual water savings from WSL. 
The total annual savings for this model are 392 gallons/m$^2$ (SE = 7.83). We believe the gap between landscape conversion and approval is substantial enough to justify the exclusion of $\tau = -1$ from the preferred model because this gap suggests that is it likely that many of the water savings that accrue between $\tau = -2$ and $-1$ are likely the result of pre-approval landscape conversions.  Model 5 therefore provides a lower bound on WSL-induced water savings. 

In the event study, we also noted that the pre-treatment decline appears to begin between years $-3$ and $-2$, so we also test a specification in which both $\tau = -1$ and $-2$ are excluded from the population (not shown). In this specification, we find that estimated water savings increase by about four percent compared to Model 3. This likely represents an overly optimistic estimate of the true WSL induced savings, and may be attributing other water conserving behavior that significantly pre-dates the landscape conversion to the WSL program.

Utilizing Model 3, we find that the seasonal pattern of WSL savings is consistent with expectations and the seasonal pattern of consumption across all models (see Fig.~\ref{fig:seasonality}). Even in winter, the estimated water savings are non-trivial, roughly 18 percent of summer conservation levels. This suggests that there is substantial outdoor water consumption even in the winter in this environment. In Model 3, our preferred specification, total savings cumulate to 438 gallons/m$^2$ (SE=8.86) annually. 

This is in comparison to the estimates generated by \citeA{sovocool_-depth_2006} of about 600 gallons per m$^2$ annually, with 103 gal/m$^2$ in July and 16.8 gal/m$^2$ in December.  Even our upper bound estimates lie well below those of Sovocool et al.

\subsection{Durability and Cohort Effects}

The results for the durability regression estimated from Eq. \ref{eq:rebound} show a very modest attrition (one to two percent of seasonal water savings per annum) in the spring summer, and winter, and slight annual increase in the fall.  The total annual decline in WSL induced savings sums to 1.94 gallons per m$^2$ per year (SE = 0.94), which is significant at the 5\% level. Taken at face value, these results imply an annual reduction in the original water savings from WSL of 4.4 percent over ten years.

However, given the structural correlation between WSL age and the timing of landscape conversion due to our time limited dataset, the potential for attrition of water savings over time must be assessed while also controlling for the effect of WSL cohort.  
When separately controlling for the effects of cohort and landscape age as in Eq. \ref{eq:rebound cohort}, the magnitude and significance of the rebound effect grows even smaller.  Annually, there is a reduction in effectiveness of 1.76 gallons per m$^2$ per year (SE = 0.92), which is not significant at the 5\% level. 

\begin{figure}[h]
   \centering
   \includegraphics[trim={0 1.5cm 0 0},clip]{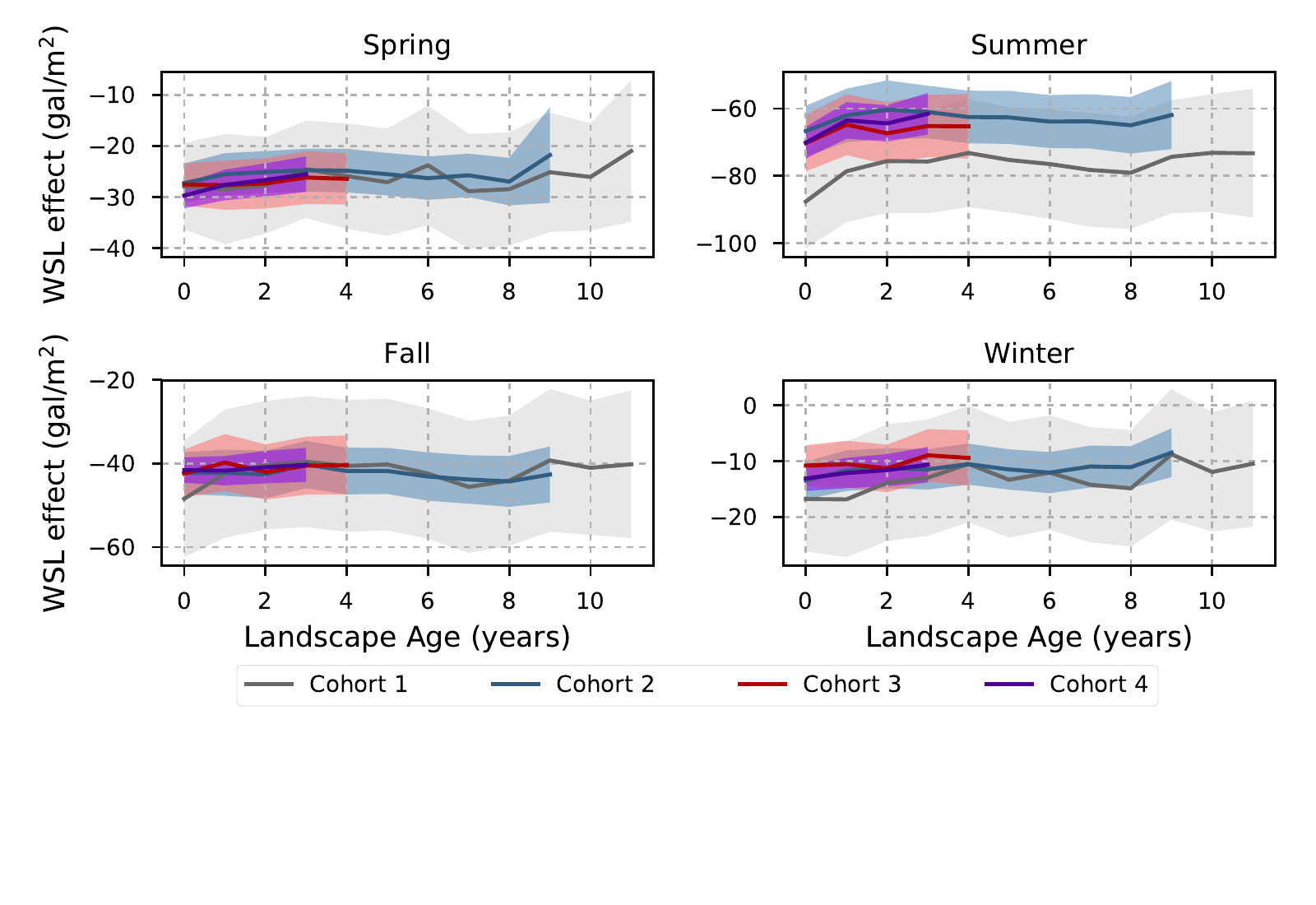} 
   \caption{Regression results showing the relationship between the age of a WSL conversion and the water savings generated. }
   \label{fig:cohort_dura_intx}
\end{figure}

Contrary to the finding of positive rebound effects in many studies of energy conservation investments, these analyses show no compelling evidence for a long-run rebound effect of WSL for water conservation in Las Vegas. 
The results of an alternate specification, in which the effect of the age of the WSL landscape is allowed to vary by year (rather than conforming to a linear slope), are shown in Fig. \ref{fig:cohort_dura_intx} and provide a less parametric confirmation of this finding.

Fig. \ref{fig:cohort_dura_intx} suggests a simple explanation for the difference between the ``Rebound''  and the ``Rebound x Cohort'' results; later adopters of WSL (who by construction have many young landscapes in our sample) appear to have conserved slightly less water per unit area upon conversion than early adopters, although the differences are small, inconsistent across seasons, and insignificant. 

\subsection{Economic Performance}

We consider the economic performance in terms of its cost effectiveness - how much water was conserved per dollar of public investment? We also consider the private economic gains, in terms of reduced water bills, to homeowners.

\subsubsection{Cost Effectiveness}

\begin{table}[h!] 
\centering
\def\sym#1{\ifmmode^{#1}\else\(^{#1}\)\fi}
\topcaption{Average gallons saved per dollar spent on rebate if we assume that the rebate expenses are annualized over a period of 5, 10, 20  or 40 years and WSL induced water savings last the same number of years. The Annual row shows the the year round average monthly water savings for each monthly dollar spent on rebates, computed as the weighted average of the four seasonal estimates.} 
\begin{tabular}{lr rrrr }
\toprule
&  & \multicolumn{4}{c}{Repayment Period}	\\
	& 	&	5 years	&	10 years	&	20 years	& 40 years	\\
	\midrule
\input{payment_1.tex}
\midrule
\input{payment_2.tex}
\rule{0pt}{3ex}
\input{payment_3.tex}
\rule{0pt}{3ex}
\input{payment_4.tex}
\bottomrule
\multicolumn{5}{l}{\footnotesize Standard errors in parentheses}\\
\multicolumn{5}{l}{\footnotesize \sym{*} \(p<0.05\), \sym{**} \(p<0.01\), \sym{***} \(p<0.001\)}\\
\label{tab:rebate_annuity}
\end{tabular}%
\end{table}

Season-specific estimates of the monthly gallons of water conserved per year-2000 dollar, $\beta$ from Eq. \ref{eq:cost effect baseline}, are shown in Tab. \ref{tab:rebate_annuity}. These estimates measure the average flow of water savings secured by the annuitized subsidy payment implied by the lump-sum subsidies to homeowners - the monthly water savings associated with an additional monthly dollar spent on WSL rebates. The estimates in different rows reflect different assumptions about the number of years of additional water savings provided by WSL, where the horizon for calculating the annuitized subsidy payment is matched to this interval. The final row of Tab. \ref{tab:rebate_annuity} presents the annual average gal/\$ estimated from a weighted sum of the four seasonal $\beta$ estimates.

The water savings per dollar vary significantly depending on assumptions about the horizon of the public investment. Under the relatively conservative assumption that WSL  secured 10 years of water savings on a typical property, we find that for every dollar spent on the WSL program, about 345 gallons of water are saved at a cost of \$2.90/kgal. In comparison, if WSL secured 20 years of additional water savings, then the water savings increases to 618 gal/\$ (\$1.62/kgal). These values straddle the retail pricing of water of about \$2.23/kgal.\footnote{This is the average price per thousand gallons paid by all consumers in the sample across all years.}

\begin{table}[ht]
\def\sym#1{\ifmmode^{#1}\else\(^{#1}\)\fi}
   \centering
   \topcaption{Average gallons saved per dollar spent on rebates assuming that WSL savings last 10 years, by cohort group. The base payment coefficient can be interpreted as the water savings for WSL conversions at age 0 in the fourth and most recent cohort.} 
   \begin{tabular}{l r r r r r}
   \toprule
\input{payment_cohort.tex}
\bottomrule
\multicolumn{5}{l}{\footnotesize Standard errors in parentheses}\\
\multicolumn{5}{l}{\footnotesize \sym{*} \(p<0.05\), \sym{**} \(p<0.01\), \sym{***} \(p<0.001\)}\\
 \end{tabular}
   \label{tab:rebate_cohort}
\end{table}

\begin{figure}[htb]
   \centering
   \includegraphics[height=5cm]{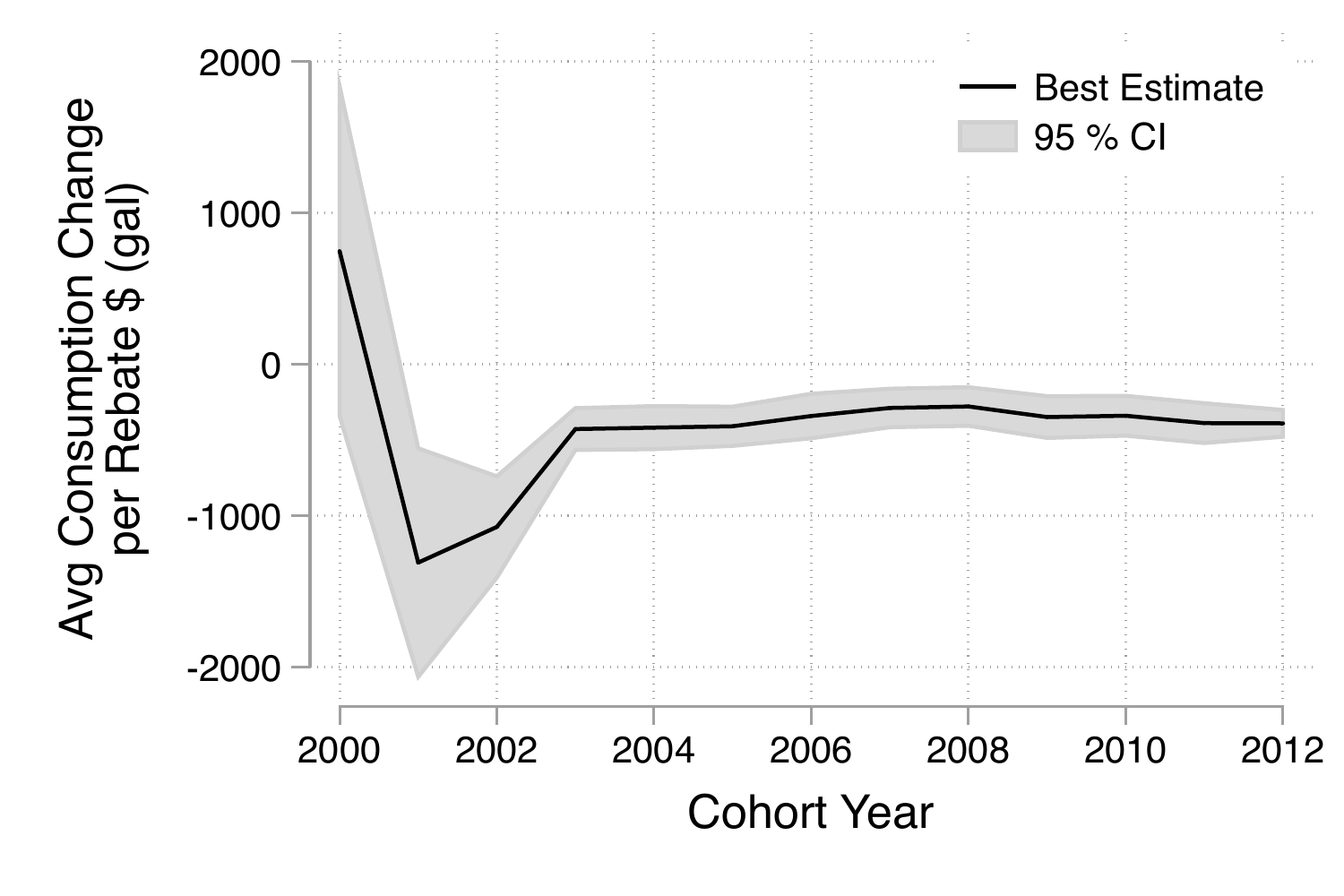} 
   \caption{Gallons of Water saved per rebate dollar invested by the year in which the WSL landscape conversion was completed. } 
   \label{fig:cost_cohort}
\end{figure}

The results estimated from Eq.~\ref{eq:cost effect cohort} are shown in Tab.~\ref{tab:rebate_cohort}. Fig.~\ref{fig:cost_cohort} shows results from an alternate specification, where the total annual gallons of water saved per dollar invested are allowed to vary with each enrollment year instead of with the cohort group. Both specifications tell a consistent story; the early cohorts, particularly Cohort 1, were more cost effective than later cohorts, largely because of the generally escalating pattern of WSL subsidies with time, but also because of the slightly greater water savings per area for the earliest cohorts.  
Cohort 3, with the highest rebate value, was the least cost effective, with about 13 percent less water saved per dollar than the most recent cohort over the course of a year (although this difference is not statistically distinguishable).  
Overall, these results suggest that cost effectiveness of the program has generally been stable after the initial phase. Finally, Tab. \ref{tab:rebate_cohort} shows that the previous finding that WSL-driven water conservation is durable is mirrored for cost-effectiveness as well.   

\subsubsection{Private Benefits}\label{sec:private benefits}

The results of the regression described in Eq.~\ref{eq:private baseline} are shown in Tab. \ref{tab:bill_base}.  The final column shows the total annual savings estimated from a weighted sum of the four seasonal regressions.  We find an annual water bill savings of 79\cent \space per m$^2$ of turf converted under WSL. Given a typical landscape conversion cost of \$15/m$^2$ \cite{sovocool_-depth_2006,snwa_water_2014}, this gives an unsubsidized and undiscounted repayment period of nineteen years. We do not consider the potential benefits homeowners may accrue from reduced costs (in dollars and time) to maintain a xeric landscape in comparison to a mesic landscape, which may reduce the repayment period. This demonstrates that the private savings on individual's water bills from turf removal were likely inadequate to have functioned as a strong inducement for homeowners to have undertaken these changes without explicit subsidization or large increases in the retail price of water. This judgement is apart from any utility gained from xeric landscaping itself; however, evidence from similar real estate markets \cite{klaiber_like_2017} suggests that xeric landscapes capitalize negatively relative to green yards, where this finding is net of any cost savings from xeric landscaping being capitalized into home prices. This reinforces the assessment that few homeowners would have undertaken turf removal without WSL subsidies. 

\begin{table}[h]
\def\sym#1{\ifmmode^{#1}\else\(^{#1}\)\fi}
   \centering
   \topcaption{This table shows the estimated private monthly savings (in year 2000 cents) for each square meter of turf converted to xeric landscaping under the WSL program. } 
   \begin{tabular}{l r r r r r}
   \toprule
\input{bill_base.tex}
\bottomrule
\multicolumn{5}{l}{\footnotesize Standard errors in parentheses}\\
\multicolumn{5}{l}{\footnotesize \sym{*} \(p<0.05\), \sym{**} \(p<0.01\), \sym{***} \(p<0.001\)}\\
 \end{tabular}
   \label{tab:bill_base}
\end{table}

Median (average) WSL conversion areas are approximately $90$ $m^2$ ($125$ $m^2$), so the median (average) annual reductions to the water bill are about \$71 (\$99). While small relative to the cost of financing a landscape conversion, these savings are nonetheless about 24 percent of the annual pre-conversion water bill for the typical WSL participant.

\section{DISCUSSION \& CONCLUSIONS}

Our DID analysis provides robust evidence that households that accepted WSL subsidies to modify their water-intensive landscaping saw substantial reductions in water use compared to households that did not take advantage of the  subsidies ($ATT^{WSL}$). As noted in Section \ref{sec:methods baseline} there are ample reasons to expect that this estimate is approximately the same as a DID comparison between households taking on WSL-style landscape transformations and those that do not ($ATT^{WSL} \approx ATT^{INSTALL}$). WSL subsidies were substantial and the WSL program was aggressively promoted such that awareness of the program was widespread by the mid-2000s. These factors, combined with the finding in Section~\ref{sec:private benefits} that the landscaping changes mandated under WSL were unattractive as private investments, suggests that few households in our non-WSL control group engaged in significant turf removal and replacement without utilizing the subsidy.  

A critical question for policy makers is the \textit{additionality} of the WSL subsidy. If all large scale landscape conversions are directly driven by the policy itself then the entire estimated average water savings of the program can be attributed to the subsidy: $ATT^{INSTALL}=ATT^{SUBSIDY}$. On the other hand, if the WSL-treated households would have removed turf from their landscapes at a higher rate than the control group in the absence of the subsidy, then $ATT^{WSL}$ will overstate subsidy-driven water savings \cite{bennear_municipal_2013}. There are several arguments suggesting a high degree of additionality for the WSL program. First, the private benefit-cost analysis for turf removal is unattractive in the absence of substantial preferences for water conservation or desert landscaping. Second, unlike many other durable goods such as refrigerators, air conditioners, or toilets \cite{Davis_cash_2014, bennear_municipal_2013} there is no clear upper bound on the useful life of landscaping. Individual plants may require maintenance or replacement, but there is no natural replacement horizon for the landscape itself. Therefore, compared to many other subsidies for the replacement of durable goods, there is little reason to suspect that WSL-rewarded landscape renovations would have occurred in the absence of the subsidy.\footnote{It is possible that some households that had already planned to undertake a major renovation of their landscaping were induced by the WSL subsidy to install a more water-conserving landscape than they would have otherwise chosen. Our DID estimates may overstate the additionality of WSL for these households if their counterfactual landscapes would have already been more water-saving than landscape changes in the control group. We have no data on the potential size of this segment or their landscaping preferences.}       

The water savings from WSL were significant, yielding reductions in annual water use of about 18 percent (48,600 gallons) for the average participant. While sizable, these reductions are smaller than estimates from an earlier pilot study \cite{sovocool_-depth_2006}, which estimate 96,000 gallon annual savings for the average participant. There are many potential causes for this gap. The Sovocool et al. study utilizes data from a pilot study that largely predates the period of the mainstreaming of WSL that we study. A combination of selection toward water-conscious early adopters and potentially more attentive calibration of irrigation equipment in the pilot period may have lead to optimistic estimates of water savings compared to under full-scale implementation.\footnote{Landscape installers may have an incentive to calibrate irrigation equipment to water more heavily than necessary in the long run to ensure the establishment of the new landscape. Inattentive homeowners may subsequently fail to adjust the watering level to an appropriate maintenance level. Landscape maintenance companies may also have an incentive to aggressively water xeric landscapes in order to ensure that their customers (who face relatively low water prices and may be inattentive to outdoor water use) are satisfied with their services.} The \citeauthor{sovocool_-depth_2006} study directly measures water application for outdoor irrigation through use of submeters.  This is clearly the ideal choice to estimate changes in outdoor water consumption, but would miss any potential offsetting behaviors if residents begin to use more water indoors, for example by taking longer showers, responding less urgently to leaks and running toilets, or running their dishwashers or washing machines more often. These manifestations of the rebound effect \cite{gillingham_energy_2013} could have appeared relatively quickly in response to lower water bills. Our analysis considers the household water use as a whole, and therefore subtracts any offsetting effects from our initial estimate of WSL-induced water savings.  Finally, it is possible that WSL indirectly induced complementary water-hungry landscape investments on the part of homeowners. For example, a homeowner may chosse to overhaul their landscaping by bundling turf removal and xeric landscaping with a new pool or water feature. The availability of WSL subsidies may have induced (or shifted forward) such bundled conversions by lowering the overall cost of the remodel. Furthermore, there may be benefits in terms of lowered costs and reduced logistical headaches from taking on these landscape investments simultaneously. However, the presence of a downward bias requires that WSL-adopters undertake these water-intensive investments at a greater rate than their neighbors in the control group. Unfortunately, the assessor data, which does not provide longitudinal information on pool and water feature installation, is incapable of assessing the magnitude of these effects. 

While much of the water savings accrued in the warm spring and summer months, we find that WSL conserved water across the entire year. This was driven in part by the relatively arid conditions that persist in Las Vegas year-round as well as by the common landscaping practice of overseeding Bermuda grass - a heat-tolerant species that goes dormant in the winter - with annual ryegrass in the fall and winter months. Indeed, the much larger water savings in fall compared to spring (Table \ref{tab:baseline}) likely comes from avoiding the multiple waterings per day required to germinate and establish a winter ryegrass lawn. 

We find little evidence of an erosion of water savings from turf removal. This persistence has multiple potential explanations. First, most xeriscaped landscapes are watered using automated timers; once these systems are calibrated (in many cases by hired landscapers) many homeowners may have a tendency to ignore the outdoor watering until a major event (e.g., a broken irrigation pipe, an excessively high water bill, or dying plants) occurs. Second, unlike many household appliances, where greater energy or water efficiency may directly induce more intensive use of the appliance over time due to the lower cost of its services (i.e., turning down the thermostat on a more efficient air conditioner), there may not be a similar intensive margin for households to exploit with respect to their landscaping. Watering more intensively need not produce additional landscape services. While there may have been significant rebound effects from WSL (see above), we expect that these developed over a short horizon after the new landscape was installed so that the initial water savings of WSL were durable after the first year of installation.         

The cost-effectiveness of WSL depends critically upon the assumed horizon of the public investment - the average length of time until a WSL-like landscape would have occurred on treated parcels in the absence of the program. This assumption is difficult to substantiate given the lack of a natural replacement horizon for landscaping. However, an investment horizon of at least 20 years seems reasonable given the durability of landscape features. In this case 1000 gallons can be conserved for \$1.62 (\$0.99 if water savings are accrued over 40 years). 

By comparison, the average annual water bill for a Las Vegas residential customer ranged between \$325 and \$395 during the study period, giving an overall average retail price of \$2.23/kgal.  The lowest marginal price charged for water - which is likely substantially below the marginal cost of supply - has declined from \$0.98 and \$0.89 during the study period, while the highest has increased from \$2.27 to \$3.56 (nominally \$4.58) over the same period.  To the extent that the average retail price approximates the marginal cost of pumping, treating and delivering water from existing supplies (primarily from Lake Mead), it suggests that the cost of reducing water use through WSL is less than the costs of supplying that same amount of water to customers.\footnote{This comparison does not account for any marginal administrative costs associated with WSL.} 

Given the scarcity and insecurity of Las Vegas' Colorado River allocation and the drought that strongly shaped Las Vegas' water policy in the 2000s, it arguable that a much more relevant comparison to the costs of water savings through WSL is the marginal cost of augmenting supplies. However, for the short to medium-term horizon for which WSL was designed, Las Vegas had (and continues to have) few means to augment its supply aside from water conservation. While some western cities have been able to expand their water supplies through purchasing agricultural water rights, Las Vegas has not been able to do so in recent decades due to a combination of limited surface-watered agriculture in southern Nevada, political and infrastructure barriers to transfers within-state, and institutional barriers to interstate transfers. With very limited surface water available locally, Las Vegas has looked to regional groundwater sources to augment supply.  In 1989, Las Vegas began applying for water permits to access groundwater from more northern parts of the state, especially the Snake Valley Aquifer, which underlies which both Nevada and Utah.  A multi-billion dollar pipeline was planned to move the water to Las Vegas.
These efforts have faced substantial opposition from ranchers and rural residents of the areas in both Nevada and Utah, and despite nearly three decades of effort and litigation, construction still has not yet started \cite{hall_interstate_2013,gehrke_utahs_2013,longson_snake_2011,jenkins_vegas_2009, green_quenching_2008}. 
Indeed, a widely-used database of water transfers in the western US from 1987 to 2009 reports \textit{no} purchases of water rights by the City of Las Vegas or the Las Vegas Valley Water District in the period of our study~\cite{Donohew_Water_2017}. 
Therefore, while we lack a concrete estimate of the cost of augmenting supply to Las Vegas, it seems clear that options for obtaining water at any price are highly uncertain, and would certainly be substantially larger than the prices charged to retail customers.  

Given the prohibitively high cost of augmenting supply in the near-term, the relevant economic context for a budget-constrained Las Vegas policy maker is how the publically borne cost of a quantity of water conservation through WSL compares to other means of saving water.\footnote{A full social benefit-cost analysis would need to include the direct costs of landscape conversion borne by homeowners, potentially offset by lower maintenance costs, as a cost of the program. Furthermore, the costs of the subsidy, while relevant to the utility, represents a transfer from the water utility to homeowners and is therefore not a social cost.} Throughout the 2000s Las Vegas pursued a multipronged policy of water conservation. In addition to stringent restrictions on turf in new construction and other construction incentives and regulations, programs targeted at existing residents including the enhanced enforcement for outdoor water waste, coupons for pool covers, rain sensors, and other irrigation systems, restrictions on the use of water features, retrofit packages for indoor fixtures in single family homes, and an award-winning publicity campaign to promote water conservation outdoors \cite{snwa_water_2009,snwa_water_2014}.

In a recent analysis of water policies in Albuquerque, NM \citeA{Price_Low-Flow_2014} estimate that cost-effectiveness of utility rebates ranged from $\$0.39/kgal$ for low-flow showerheads, $~\$1/kgal$ for dishwashers and washing machines, and over $\$8.00/kgal$ for the replacement of low-flow toilets.\footnote{Costs of water savings in this and other papers we report utilize a variety of, often unspecified, assumptions on the use of nominal vs. real prices, discount rates, and the method used to attribute water savings to program costs. We do not attempt to resolve these differences; therefore comparisons should be made cautiously.}\textsuperscript{,}\footnote{They also find that a xeriscape rebate program cost $\$4.51/kgal$. The greater cost-effectiveness of the Las Vegas program may have been driven in large part by the greater potential year-round water savings from turf removal in Las Vegas relative to Albuquerque.} However, these calculations rely upon a common but strong assumption - that all subsidized appliance replacement is additional. Yet, there are longstanding concerns that many participants in water- and energy-efficiency programs are free riders that would have undertaken the desired behavior in the absence of the subsidy \cite{joskow_what_1992}. \citeA{bennear_municipal_2013} utilize data from Cary, NC to estimate that over 67 percent of the water savings associated with high-efficiency toilet rebates would have occurred without the rebates, increasing the cost of water savings to $\$10.85/kgal$ if the lifespan of existing toilets was 15 years. \citeA{Boomhower_credible_2014} estimate that approximately half of individuals purchasing new energy-efficient refrigerators and appliances under a Mexican subsidy program were non-additional. This suggests that subsidies for replacement of appliances and fixtures may be considerably less cost-effective than commonly presumed. 

An alternative approach to pecuniary incentives is to utilize informational campaigns and nudges rooted in pro-social norms to alter household behavior directly. This approach is now being mainstreamed through customer engagement programs for utilities such as WaterSmart Software and Opower. \citeA{Ferraro_Using_2013} demonstrated that programs that go beyond information provision by comparing individuals' water use to their neighbors' can be highly cost-effective, reducing water use by nearly 5 percent at a cost of $\$0.58/kgal$. However, the ability of these behavioral interventions to provide sustained water savings remains controversial. \citeA{Ferraro_Using_2013} found that effects attenuate quickly, yet in a follow-up study \citeA{Bernedo_persistent_2014} report that effects remain policy-relevant six years later - reducing costs of water conservation to $\$0.24/kgal$. \citeA{Allcott_short-run_2014} suggest that repeated exposure to socially framed information provision on energy use may slow the rate of backsliding - yielding long-run conservation effects that decay relatively slowly. Indeed, while Las Vegas did not engage in targeted behavioral nudges, they did nonetheless utilize mail and television marketing to promote drought awareness and water conservation behaviors. \citeA{Brelsford_growing_2017} provide suggestive evidence that these efforts may have played a substantial role in explaining the large reductions in Las Vegas' per-capita water use in the mid-2000s. 

Examining the wide range of cost-effectiveness estimates suggests that WSL compares favorably to many rebate programs, yet perhaps less so compared to informational/nudge-based programs. While our estimates suggest that WSL has not fully lived up to the the optimistic water savings and cost-effectiveness calculations of early pilot studies \cite{sovocool_-depth_2006}, it nevertheless has a number of attractive characteristics that have made it a vital part of Las Vegas' water policy toolbox. Its effects on individual water conservation have been demonstrably large (approximate 18\% on average), while, at its best, norm-based messaging reduces water use by 5\%.  Reducing \textit{outdoor} water use was especially important given that much of the water used outdoors does not return to Lake Mead and cannot be credited against Las Vegas' allocation of the Colorado River through return flow credits. As a result, WSL provided a cost-effective pathway to permanently augment Las Vegas' water supply through water conservation at a time when the city was beset by a severe drought and when alternative sources of supply were not readily available. Whether similar landscape subsidy programs would be as attractive in other jurisdictions is unclear; Las Vegas' highly arid climate may enhance the cost-effectiveness of WSL relative to other subsidy programs. Nevertheless, our results suggest that subsidies for turf removal may be a promising way for budget-constrained utilities facing high costs of tapping new supplies to effectively enhance their existing supply by building future water efficiency into the urban landscape.        

\clearpage
\bibliography{Vegas}
\bibliographystyle{apacite}

\end{document}

%% file: baseline_1.tex
WSL Area&Spring&-27.28\sym{***}&-26.78\sym{***}&-26.44\sym{***}&-26.73\sym{***}&-23.09\sym{***}\\
&&(0.97)&(1.11)&(0.95)&(1.10)&(0.81)\\
&Summer&-66.19\sym{***}&-64.89\sym{***}&-63.92\sym{***}&-64.80\sym{***}&-58.48\sym{***}\\
&&(1.99)&(2.32)&(1.95)&(2.30)&(1.75)\\
&Fall&-43.85\sym{***}&-42.62\sym{***}&-41.76\sym{***}&-42.54\sym{***}&-36.70\sym{***}\\
&&(1.45)&(1.69)&(1.42)&(1.68)&(1.24)\\
&Winter&-12.20\sym{***}&-11.89\sym{***}&-11.56\sym{***}&-11.87\sym{***}&-9.69\sym{***}\\
&&(0.62)&(0.72)&(0.61)&(0.72)&(0.54)\\

%% file: baseline_2.tex
R$^2$&Spring&0.137&0.195&0.122&0.122&0.115\\
&Summer&0.277&0.376&0.246&0.246&0.236\\
&Fall&0.194&0.268&0.173&0.174&0.165\\
&Winter&0.075&0.105&0.071&0.071&0.069\\

%% file: baseline_3.tex
Households&&52,571&26,285&52,571&52,571&52,571\\

%% file: baseline_4.tex
Observations&Spring&590,040&303,269&632,272&632,272&658,443\\
&Summer&588,903&302,304&630,033&630,033&655,983\\
&Fall&596,345&305,046&635,775&635,775&661,792\\
&Winter&598,444&305,606&637,165&637,165&663,143\\

%% file: payment_1.tex
Payment&Spring&-133.65\sym{***}&-252.59\sym{***}&-452.64\sym{***}&-736.57\sym{***}\\
&&(4.52)&(8.54)&(15.30)&(24.89)\\
&Summer&-320.40\sym{***}&-605.55\sym{***}&-1,085.15\sym{***}&-1,765.85\sym{***}\\
&&(9.26)&(17.51)&(31.38)&(51.06)\\
&Fall&-206.67\sym{***}&-390.59\sym{***}&-699.95\sym{***}&-1,139.02\sym{***}\\
&&(6.79)&(12.84)&(23.01)&(37.44)\\
&Winter&-56.94\sym{***}&-107.61\sym{***}&-192.83\sym{***}&-313.79\sym{***}\\
&&(2.88)&(5.45)&(9.76)&(15.89)\\
&Annual&-182.50\sym{***}&-344.91\sym{***}&-618.09\sym{***}&-1,005.81\sym{***}\\
&&(3.51)&(6.63)&(11.88)&(19.34)\\

%% file: payment_2.tex
R$^2$&Spring&0.120&0.120&0.120&0.120\\
&Summer&0.242&0.242&0.242&0.242\\
&Fall&0.171&0.171&0.171&0.171\\
&Winter&0.071&0.071&0.071&0.071\\

%% file: payment_3.tex
Households&&52,571&52,571&52,571&52,571\\

%% file: payment_4.tex
Observations&Spring&632,272&632,272&632,272&632,272\\
&Summer&630,033&630,033&630,033&630,033\\
&Fall&635,775&635,775&635,775&635,775\\
&Winter&637,165&637,165&637,165&637,165\\

%% file: payment_cohort.tex
&\multicolumn{1}{c}{Spring}&\multicolumn{1}{c}{Summer}&\multicolumn{1}{c}{Fall}&\multicolumn{1}{c}{Winter}\\
\hline
Payment             &      -233.9\sym{***}&      -548.3\sym{***}&      -345.5\sym{***}&      -104.6\sym{***}\\
                    &      (9.04)         &      (18.7)         &      (13.3)         &      (7.00)         \\
Cohort 1 $\times$ Payment&      -530.1\sym{***}&     -1467.9\sym{***}&      -753.7\sym{***}&      -259.8\sym{*}  \\
                    &      (82.1)         &     (169.4)         &     (168.6)         &     (103.4)         \\
Cohort 2 $\times$ Payment&       -63.1\sym{***}&      -168.8\sym{***}&      -126.1\sym{***}&       -36.7\sym{***}\\
                    &      (13.7)         &      (23.9)         &      (22.4)         &      (10.8)         \\
Cohort 3 $\times$ Payment&        20.8         &        31.9         &        33.0         &        19.3\sym{*}  \\
                    &      (11.4)         &      (23.3)         &      (17.8)         &      (9.09)         \\
Age $\times$ Payment&        2.15         &        2.37         &       -1.29         &        2.94\sym{**} \\
                    &      (1.12)         &      (2.05)         &      (1.69)         &      (0.90)         \\
\hline
\(R^{2}\)           &       0.121         &       0.245         &       0.173         &       0.071         \\
Households          &      52,571         &      52,571         &      52,571         &      52,571         \\
Observations        &     632,272         &     630,033         &     635,775         &     637,165         \\

%% file: bill_base.tex
&\multicolumn{1}{c}{Spring}&\multicolumn{1}{c}{Summer}&\multicolumn{1}{c}{Fall}&\multicolumn{1}{c}{Winter}  &\multicolumn{1}{c}{Annual} \\
\hline
WSL Area            &       -4.53\sym{***}&       -12.0\sym{***}&       -7.64\sym{***}&       -1.67\sym{***}  & -79.02\sym{***} \\
                   &      (0.23)         &      (0.54)         &      (0.36)         &      (0.14)       &   (2.39) \\
\hline
\(R^{2}\)           &       0.032         &       0.069         &       0.037         &       0.020         \\
Households          &      52,571         &      52,571         &      52,571         &      52,571         \\
Observations        &     632,272         &     630,033         &     635,775         &     637,165         \\